%% file: amore2_background.tex
\documentclass[twocolumn,epjc3]{svjour3}  
\smartqed  
\RequirePackage{graphicx}
\usepackage[utf8]{inputenc}
\usepackage{graphics}
\usepackage{epstopdf}
\usepackage{xcolor}
\usepackage{rotating}
\usepackage{dblfloatfix} 
\usepackage{microtype} 

\usepackage{amsmath,amssymb}
\usepackage[hypertexnames,setpagesize,%
    pdftex,%
    colorlinks,%
    citecolor=blue,%
    hyperindex,%
    plainpages=false,%
    bookmarksopen,%
    bookmarksnumbered%
  ]{hyperref} 
  
\usepackage[numbers,square,comma,sort&compress]{natbib}

\usepackage{textcomp}

\usepackage{lineno,xcolor}
\usepackage{tikz}
\usetikzlibrary{arrows,shapes,positioning}

\usepackage{orcidlink}

\usepackage{multirow}

\journalname{Eur. Phys. J. C}

\begin{document}


\title{
Projected background and sensitivity of AMoRE-II
}

\input{author-orcid-EPJC.tex}
%



\date{Received: date / Accepted: date}

\maketitle

\begin{abstract}
AMoRE-II aims to search for neutrinoless double beta decay  ($0\nu\beta\beta$) with an array of 423 Li$_2$$^{100}$MoO$_4$ crystals operating in the cryogenic system as the main phase of the Advanced Molybdenum-based Rare process Experiment (AMoRE). AMoRE has been planned to operate in three phases: AMoRE-pilot, AMoRE-I, and AMoRE-II. AMoRE-II is currently being installed at the Yemi Underground Laboratory, located approximately 1000 meters deep in Jeongseon, Korea. 
The goal of the experiment is to reach an exclusion half-life sensitivity to the $0\nu\beta\beta$ of $^{100}$Mo on the level of $\mathnormal{T}$$^{0\nu\beta\beta}_{1/2}$ $>$~6~$\times$~10$^{26}$ yr that covers completely the inverted Majorana neutrino mass hierarchy region of (15--46)~meV.
To achieve this, the background level of the experimental configurations and possible background sources of gamma and beta events should be well understood. We have intensively performed Monte Carlo simulations using the GEANT4 toolkit in all the experimental configurations with potential sources. We report the estimated background level that meets the 10$^{-4}$~counts/(keV$\cdot$kg$\cdot$yr) requirement for AMoRE-II in the Region Of Interest (ROI) and show the projected half-life sensitivity based on the simulation study.
\end{abstract}

%
	


\input{1_introduction.tex}
\input{2_experiment.tex}
\input{3_simulation.tex}
\input{4_analysis-results.tex}
\input{5_sensitivity.tex}
\input{6_conclusion.tex}

\section{Acknowledgement} 
This research was supported by the Institute for Basic Science (Korea) under project codes IBS-R016-D1 and IBS-R016-A2.
It is also supported by the Ministry of Science and Higher Education of the Russian Federation (N121031700314-5), the MEPhI Program Priority 2030.
The group from the Institute for Nuclear Research (Kyiv, Ukraine) was supported in part by the National Research Foundation of Ukraine under Grant No. 2023.03/0213. These acknowledgments are not to be interpreted as an endorsement of any statement made by any of our institutes, funding agencies, governments, or their representatives.
\bibliography{ref_simulation.bib}

\end{document}

%% file: author-orcid-EPJC.tex
\author{ 
{A.~Agrawal\,\orcidlink{0000-0001-7740-5637}}
\and {V.V.~Alenkov\,\orcidlink{0009-0008-8839-0010}}
\and {P.~Aryal\,\orcidlink{0000-0003-4955-6838}}
\and {J.~Beyer\,\orcidlink{0000-0001-9343-0728}}
\and {B.~Bhandari\,\orcidlink{0009-0009-7710-6202}}
\and {R.S.~Boiko\,\orcidlink{0000-0001-7017-8793}}
\and {K.~Boonin\,\orcidlink{0000-0003-4757-7926}}
\and {O.~Buzanov\,\orcidlink{0000-0002-7532-5710}}
\and {C.R.~Byeon\,\orcidlink{0009-0002-6567-5925}}
\and {N.~Chanthima\,\orcidlink{0009-0003-7774-8367}
% ...
} \\
(AMoRE Collaboration)
}
\thankstext{corrauthor1}{e-mail: ejjeon@ibs.re.kr}
\thankstext{corrauthor2}{e-mail: jeewon.seo.ibs@gmail.com}
%

%% file: 1_introduction.tex
\section{Introduction}
\label{sec:intro}
The discovery of neutrino oscillations established the existence of massive neutrinos, which is in contrast with the Standard Model prediction, and it commands to search the neutrinoless double beta (0$\nu\beta\beta$) decay experimentally to confirm the Majorana nature of neutrinos mediated by non-standard model mechanisms~\cite{Rodejohann2011,Giunti2015}. 

The inverse of the half-life of \( 0\nu\beta\beta \) decay, assuming the exchange of light neutrinos mediates it, is proportional to the square of the effective Majorana neutrino mass, $\langle m_{\beta\beta} \rangle^2$: 
\begin{align}
	\left(T_{1/2}^{0\nu}\right)^{-1}	&=
	 	\begin{aligned}[t]
		&G^{0\nu} \cdot |M^{0\nu}|^2 \cdot 
        \frac{\langle m_{\beta\beta} \rangle^2}{m_e^2},
		\end{aligned}
\end{align}		
where \( T_{1/2}^{0\nu} \) is the half-life of the \( 0\nu\beta\beta \) decay, \( G^{0\nu} \) is the phase space factor, \( M^{0\nu} \) is the nuclear matrix element, \( m_{\beta\beta} \) is the effective Majorana neutrino mass, and \( m_e \) is the electron mass~\cite{Mohapatra2007,Giunti2007,Agostini2023,Workman2022}.
The effective Majorana neutrino mass $\langle m_{\beta\beta} \rangle$ is a function of the masses and mixing angles of the three neutrinos, as well as the unknown Majorana phases:
\begin{align}
    \langle m_{\beta\beta} \rangle    &=
        \begin{aligned}[t]
        &\left| \sum_{i=1}^{3} U_{ei}^2 \, m_i \right|,
        \end{aligned}
\end{align}
where \( m_i \) are the neutrino masses, and \( U_{ei} \) are the elements of the Pontecorvo–Maki–Nakagawa–Sakata matrix. 
Therefore, the effective Majorana neutrino mass can be obtained from the measured half-life of 0$\nu\beta\beta$ decay.
The most sensitive lower limits on half-lives of 0$\nu\beta\beta$ decay for different isotopes, such as $^{76}$Ge, $^{82}$Se,  $^{100}$Mo, $^{130}$Te, and $^{136}$Xe are at $10^{24}$ to $10^{26}$ years~\cite{Abe2023,Agostini2020,Arnquist2023,Anton2019,Adams2020,Azzolini2022,Arnold2015,Augier2022}.

AMoRE is an experiment searching for 0$\nu\beta\beta$ decay using $^{100}$Mo isotope in molybdate crystal scintillators operated as cryogenic detectors~\cite{Bhang2012,GBKim2015}.
It has been planned to operate in three phases: AMoRE-pilot, AMoRE-I, and AMoRE-II. AMoRE-II is the main phase of AMoRE and is currently being installed at the Yemi Underground Laboratory (Yemilab), located approximately 1000 meters deep in Jeongseon, Korea. 
It uses a 180~kg array of Li$_2$$^{100}$MoO$_4$ (LMO) crystals, aiming at improving the sensitivity to the half-life of 0$\nu\beta\beta$ decay of $^{100}$Mo up to $\mathnormal{T}$$^{0\nu\beta\beta}_{1/2}$~$\sim$~6~$\times$~10$^{26}$ years. 
The sensitivity for $\mathnormal{T}$$^{0\nu\beta\beta}_{1/2}$ increases linearly with the experiment exposure if the zero background level in
the region of interest (ROI) is achieved.  
Therefore, the requirements on the background are very severe, and we set AMoRE-II requirements for the background level to be below 10$^{-4}$~counts/(keV$\cdot$kg$\cdot$yr) (ckky) in total. Monte Carlo simulations using the GEANT4 toolkit~\cite{Agostinelli2003} are conducted to assess the background level from all known possible sources.

We evaluate the background contributions from radioisotopes in the $^{238}$U, $^{232}$Th, $^{40}$K, and $^{235}$U decay chains in detectors, materials in the nearby detector system, shielding materials, and the rock walls surrounding the experimental enclosure. Additionally, we estimate the neutron and muon-induced background levels on the AMoRE-II shielding configuration at the Yemilab.
We discuss the projected half-life sensitivity and we give an outlook based on simulated background spectra and rates.

%% file: 2_experiment.tex
\section{AMoRE-II experiment}
\label{sec:experiment}
\begin{figure}[t]
\centering
\includegraphics[width=0.48\textwidth]{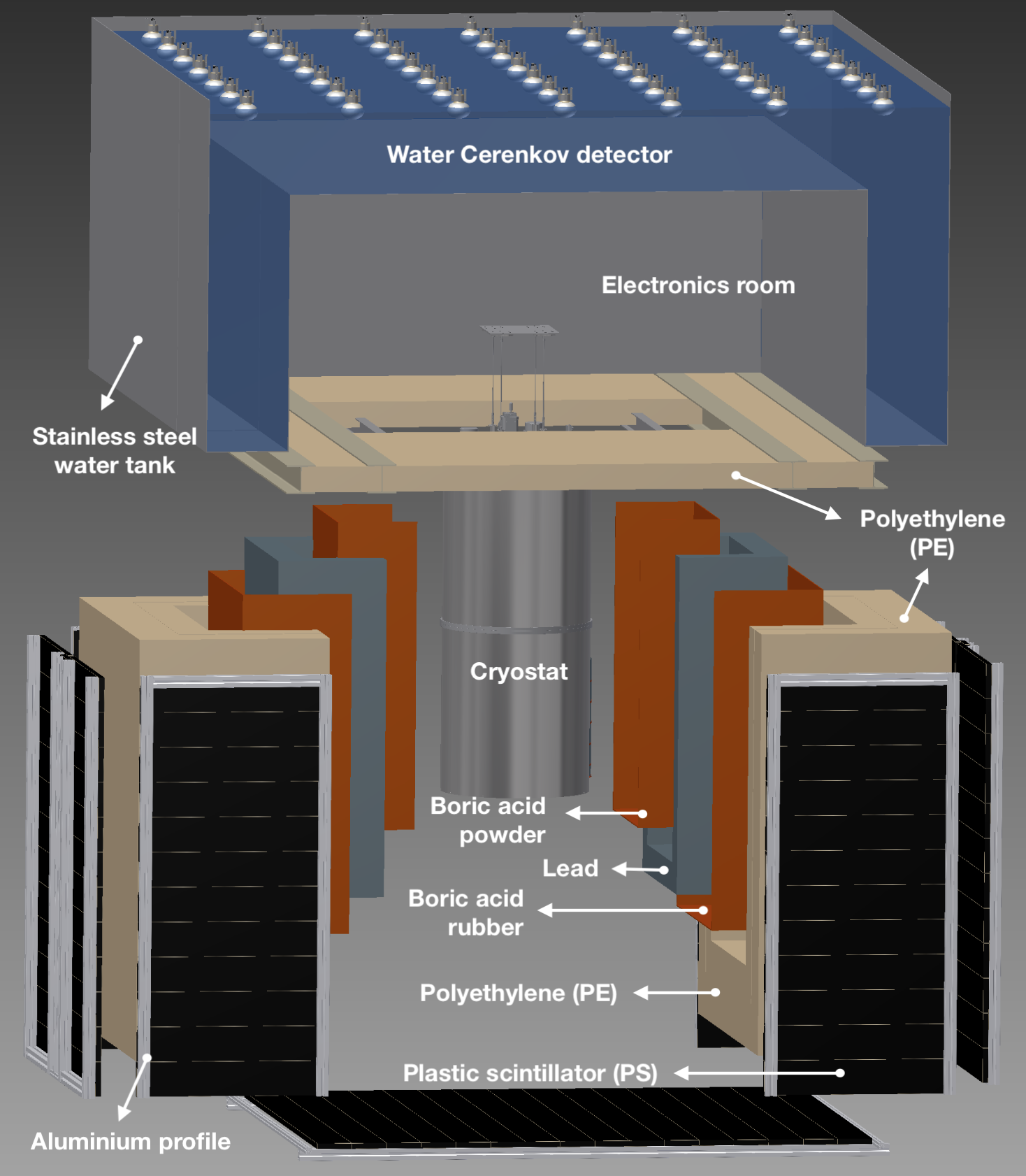}
\caption{An overview of AMoRE-II shielding and muon detector system.}
\label{fig:geo1}
\end{figure}

The AMoRE-II experiment is located at the Yemilab, approximately 1000 meters deep under Mt. Yemi in Jeongseon, Korea.
The Yemilab is a newly constructed underground laboratory that involved the excavation of a 782-meter-long tunnel from the 627-meter-long vertical shaft of the Handuk iron mine under Mt. Yemi~\cite{Yemilab2024}. The construction of the laboratory was completed in September 2022.
The AMoRE experimental hall, the second largest laboratory of the Yemilab, is a hexahedron-shaped structure that measures 21 meters in width, 21 meters in length, and 16 meters in height. The preparations for the AMoRE-II experiment are currently underway as a follow-up to the AMoRE-I experiment, which aimed to measure the neutrinoless double beta decay at Yangyang Underground Laboratory (Y2L). 
The AMoRE-II experiment uses a 423 Li$_2$$^{100}$MoO$_4$ crystals array operating at cryogenic temperatures. 
As shown in Fig.~\ref{fig:geo1}, the cryostat is surrounded by four shielding layers. The shielding construction is now complete, including plastic scintillator muon veto counters depicted in black colored panels and a water tank that includes photomultiplier tubes (PMTs) installed on the tank ceiling.
\begin{figure*}[ht]
\centering
\begin{tabular}{ccc}
\multirow{2}{*}[0.2\textwidth]{(a) \includegraphics[width=0.33\textwidth]{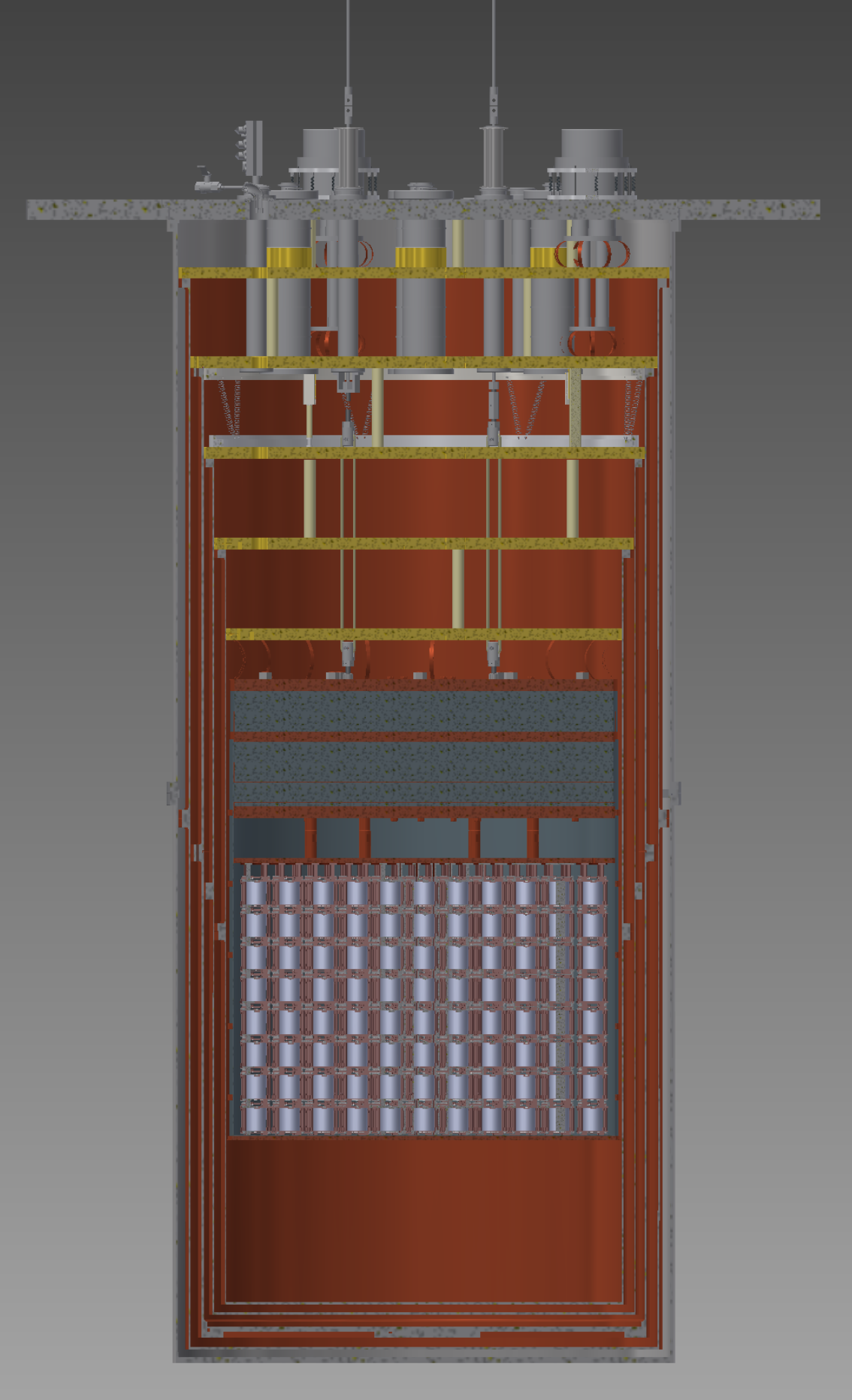}} & 
(b) & \includegraphics[width=0.25\textwidth]{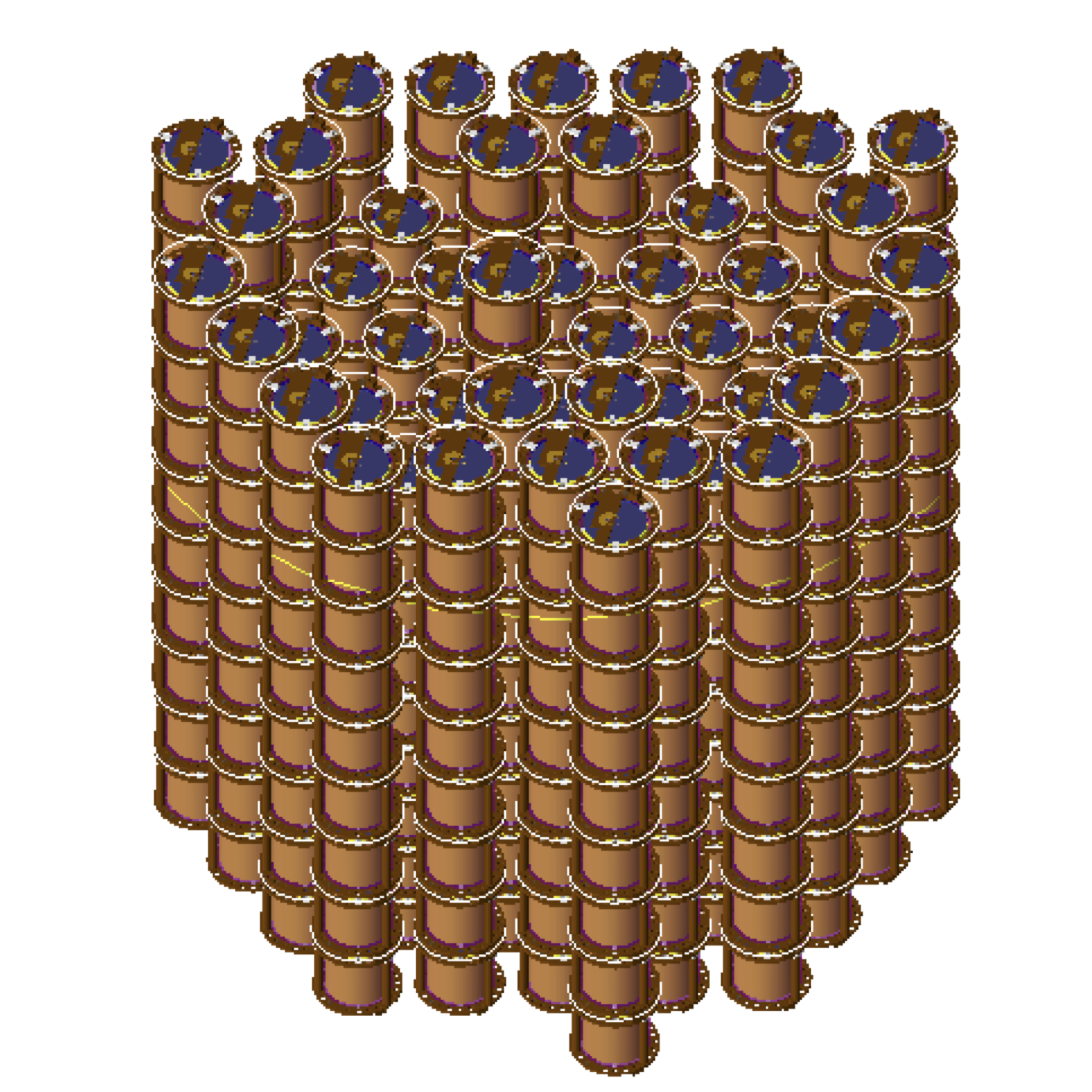}\\
& (c) & \includegraphics[width=0.35\textwidth]{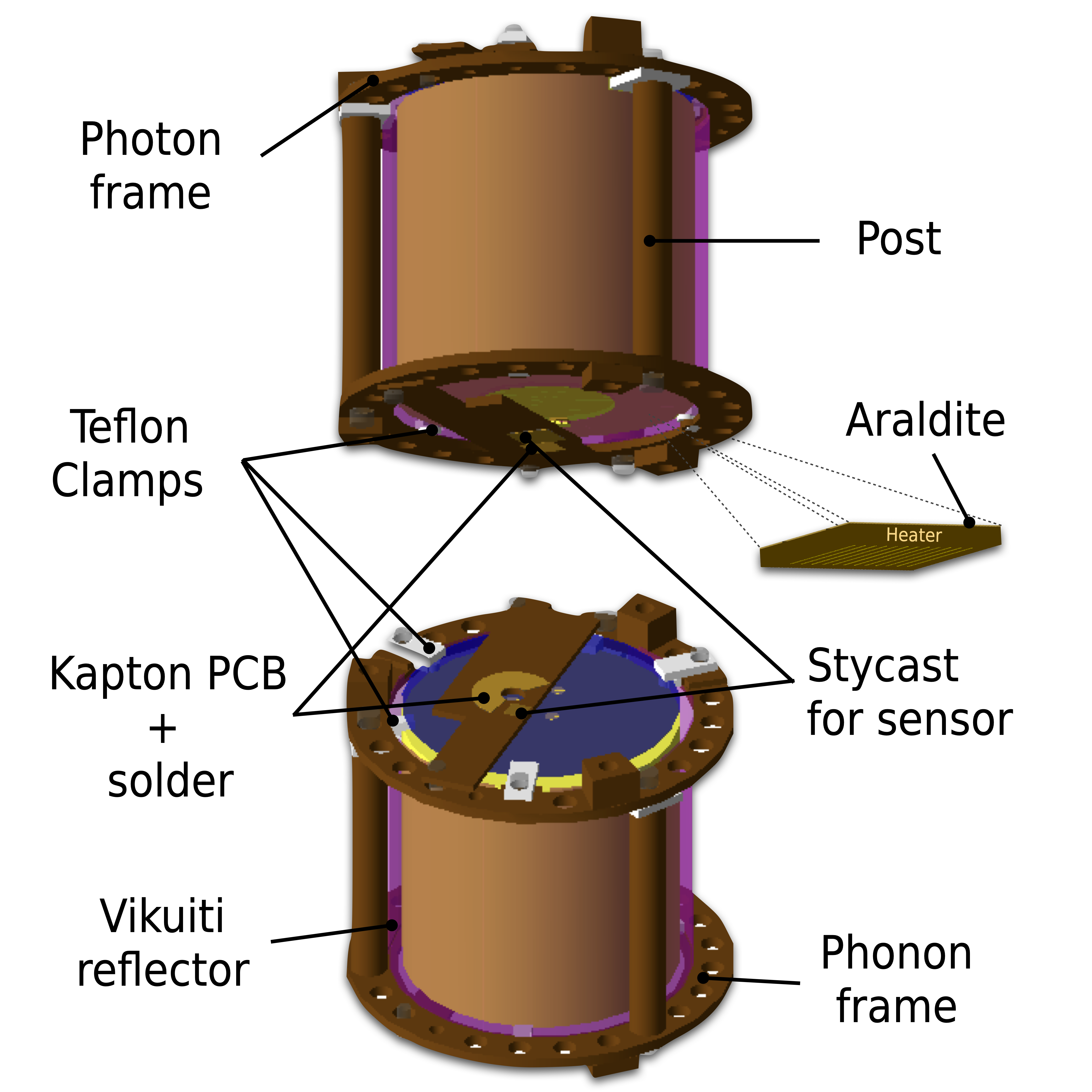}\\
\end{tabular}
\caption{
The detector geometry is shown in clockwise order: (a) Cross-sectional view of the cryostat, (b) Crystal towers, (c) Crystal detector modules.}
\label{fig:geo2}
\end{figure*}
Below, there is a detailed description of the experimental setup and detector geometry used in the simulations.
\begin{itemize}
  \item[$\bullet$] \textbf{Muon detector system} \\
  To determine the appropriate thickness for the water tank, we conducted simulation studies~\cite{Hanwook2020, Jeewon2022} and determined that 70~cm would provide a sufficient shielding. However, to accommodate the PMTs on the ceiling, we increased the thickness of the top structure by 10~cm, resulting in a final thickness of 80~cm for the top structure while maintaining the side structure at 70~cm.
The tank contains approximately 60 tonnes of water and is equipped with 48 PMTs measuring 8 inches in the inner region and 10 inches in the outer region where the water depth is greater. 
It serves as a shield and an active muon veto detector for the upper part above the cryogenic detector system and shielding layers. The lower part is surrounded by 126 veto counter modules installed like roof tiles overlapping, each consisting of 3 cm thick plastic scintillator panels (PS), as shown in Fig.~\ref{fig:geo1}. \\
  
  \item[$\bullet$]  \textbf{Shielding structure} \\
  Based on simulated design studies~\cite{Hanwook2020}, we selected a shielding configuration from outside to inward with 70-cm-thick Polyethylene (PE), 1-cm-thick boric acid rubber, 25-cm-thick lead, and 1-cm-thick boric acid powder, as shown in Fig.~\ref{fig:geo1}. Boric acid rubber is also installed inside the water Cherenkov detector to prevent the background from thermalized neutrons. 
  A total of $\sim$65 tonnes of lead were used for the 25-cm-thick lead shielding. The inner layer of the shielding, which is 5 cm thick out of 25 cm, was made of Boliden lead, which is low-radioactivity modern lead. The other 20~cm lead shield is made of normal lead produced by Nuclear Light Industry Co. Ltd. \\

  \item[$\bullet$]  \textbf{Cryostat} \\
  In Fig.~\ref{fig:geo2}(a), we present a cross-sectional view of the cryostat. 
  The cryogenic system is installed inside the shielding structure and it consists of five layers of cryostat cans. 
  The layers are arranged from the outside to the inside in the following order: a 5-mm-thick layer made of stainless steel that weighs approximately 600~kg that serves as the outer vacuum chamber (OVC) of the cryostat and four copper shielding layers (50~K, 4~K, 1~K, and 100~mK chambers) with a total thickness of 18~mm.
  Inside the copper can, the disk-shaped 26-cm-thick low-radioactivity lead structure, the inner lead, is designed to shield the background from entering from the upper part of the detector. It is supported by 203~kg of copper plates, made of annealed oxygen-free electronic (OFE) copper, not to generate backgrounds that can directly influence the crystal tower.
  Additionally, a 1-mm-thick superconducting shield layer made of LemerPax lead, an ancient Roman lead with low radioactivity, surrounds crystal detector towers below the inner lead. \\ 
  
   \item[$\bullet$]  \textbf{Crystal detector towers and modules} \\
   Inside the cryostat, as shown in Fig.~\ref{fig:geo2}(b), the crystal detectors consist of 423 modules that are arranged in 47 towers, with each tower comprising nine crystals stacked together. Each crystal is surrounded by a 65-$\mu$m-thick Vikuiti enhanced specular reflector film (VM2000)~\cite{VM2000, GILEVA2023110673} and is assembled into a module using NOSV-grade copper frames with high thermal conductivity and low radioactivity, as shown in Fig.~\ref{fig:geo2}(c). 
   The cylindrical crystals are grouped into two sizes: 50~mm diameter for 171 crystals and 60~mm diameter for 252 crystals. The corresponding modules have a diameter of 74~mm and a height of 64.7~mm (84~mm in diameter and 74.7~mm in height).
   A gold film with a diameter of 1.4 cm and a thickness of 300 nanometers is evaporated on the bottom surface of the crystal. It serves as a phonon collector thermally connected to a metallic magnetic calorimeter (MMC)~\cite{Enss2000}. The MMC measures the rise in the crystal temperature caused by radiation absorption~\cite{GBKim2017,IKim2017,YHKim2022,HKim2023}.
   A detachable photon detector is installed at the top of the copper frame~\cite{HJLee2015,MBKim2023}. It consists of a silicon oxide wafer with a diameter equal to that of the crystal and a thickness of 300 micrometers. This wafer is used as a scintillation light absorber.
   Each crystal has a stabilization heater on its bottom surface to examine the stability correction with heater pulse~\cite{DHKwon2020}. 
The Araldite adhesive attaches the heater to the crystal surface, and its background contribution is considered in the simulation.
The detector module's wiring system is designed with a Kapton-based flexible printed circuit board (PCB)~\cite{PCB}, and the soldering joints are made with pure lead-tin alloy. Although their masses are small, these components are situated near the crystals, and their impact on the background cannot be overlooked. Thus, it is factored into the simulation.
\end{itemize}

%% file: 3_simulation.tex
\section{Background simulations}
\subsection{Geant4-based Monte Carlo simulation}
We used a Geant4-based simulation framework developed for modeling background spectra of the AMoRE-Pilot and AMoRE-I experiments, which adopted G4 version 10.4.2, implemented AMoRE-specific physics lists for both internal and external background simulations; the Physics list classes of G4EmLivermorePhysics for low energy electromagnetic process, G4RadioactiveDecay for radioactive decay process, QGSP\_BERT\_HP for high-energy physics process (above 10 GeV), and the precision of the neutron model for neutrons with energies below 20 MeV were used~\cite{Agostinelli2003,Allison2006,Allison2016}. In addition, we used a full-elastic-scattering dataset for thermal neutrons with energies below 4 eV to precisely describe the shielding effect.
Using the simulation framework, we perform simulations for the background contributions from the radioactive decay chains of $^{238}$U, $^{232}$Th, $^{235}$U, and $^{40}$K.

We use the energy deposits in the crystals within a 5 ms window to simulate an event like that in the experimental data, considering the 1--2~ms rise times for crystal detectors used in the AMoRE-I experiment~\cite{Hanbeom2022}. To simulate pileup events resulting from decays with short half-lives, followed by subsequent daughter decay within a few times the typical pulse width ($\sim$20--30~ms) in cryogenic measurements~\cite{Hanbeom2022}, a time window of 100~ms is used. If the second decay happens within 5--100 ms of the first decay, it is excluded from the MC data, similar to the experimental data.
In addition, pileup events can occur due to random coincidences between sources that contribute to the ROI. The LMO crystal is a major source of background events, particularly the random coincidence of background events from the two-neutrino double beta (2$\nu\beta\beta$) decay of $^{100}$Mo. 
We also consider its impact on the background contribution to the ROI. More details about this can be found in the following section.

The energy distribution of simulated events was reconstructed by randomly smearing it in a Gaussian shape based on the resolution function obtained from the AMoRE-II R\&D setup~\cite{WTKim2022}. 

\subsection{Cosmic muons and muon-induced background}
Background caused by cosmic-ray muons is one of the most dangerous external radiation sources.
To shield those external radiations, the AMoRE-II detector is positioned around 1000 meters deep under Yemi Mountain, which corresponds to $\sim$2700 meters water equivalent (m.w.e.), and is surrounded by robust shielding materials, as described in Sect.~\ref{sec:experiment}.

To investigate the impact of cosmic muons and the induced backgrounds resulting from their interaction with the rocky cavern, shield, and detector materials, we conducted simulations using the muon energy spectrum at the Yemilab underground. This spectrum was obtained by digitizing the contour map of the Mt. Yemi area, as detailed in Ref.~\cite{Hanwook2020}. 
Due to various ambiguities from rock properties, different depths, etc., we used the total muon flux at the Yemilab for normalization, which is considered to be 8.2$\times$10$^{-8}$ muons/cm$^{2}$/s. This value was derived by considering the measured flux at Y2L~\cite{Prihtiadi2018muon} and the fact that the Yemilab is approximately 1.5 times deeper than Y2L. The recent measurement at the Yemilab of around 10$^{-7}$ muons/cm$^{2}$/s is consistent with this value; it represents a reduction by a factor of about 10$^{5}$ compared to the sea level. 

To account for all secondaries induced by muons traveling through the rocky cavern, we generated muons with the energy spectrum at the Yemilab from outside a 3-meter-thick rock shell surrounding the cavern.
The thickness of the rock shell was optimized through additional simulations. We simulated muons with an energy of 236~GeV, the mean of the muon spectrum at the Yemilab, entering the rock with varying thicknesses from 0.5 to 10 meters. We measured the mean energy and event rates of neutron and gamma secondaries emerging from the rock as a function of thickness and determined the optimal thickness, 3 meters, at the points where the function saturated.

\begin{figure}[t]
\centering
\includegraphics[width=0.5\textwidth]{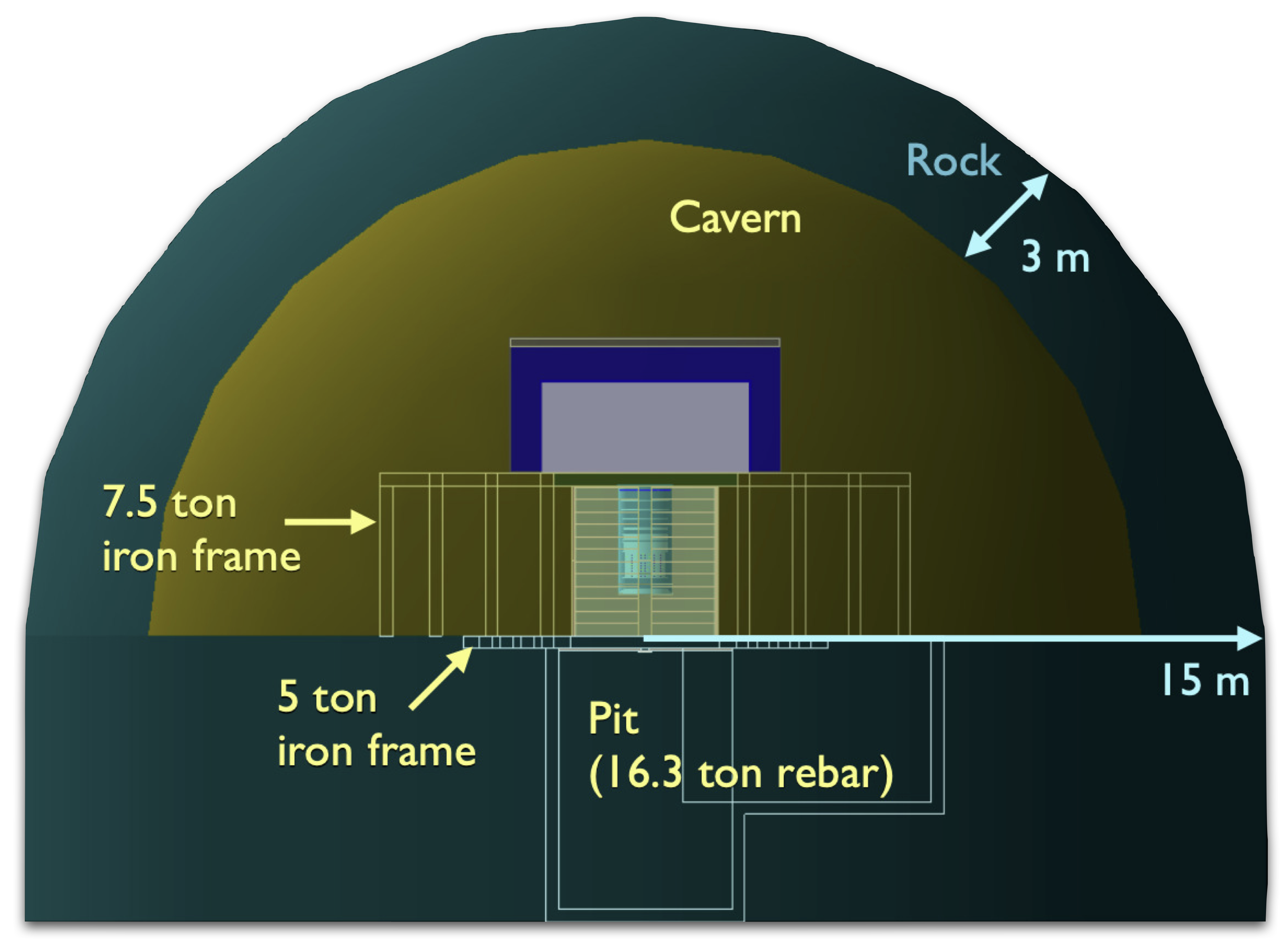}
\caption{The overall view of the geometry implemented in the simulation for the AMoRE-II experiment.}
\label{fig:OverallGeo}
\end{figure}

%
In order to estimate background rates for AMoRE-II, we analyzed the neutrons and gamma rays generated by muon interactions with materials in rock, shield, and detector components. 
The schematic view of the simulation geometry is presented in Fig.~\ref{fig:OverallGeo}.
Our study involved evaluating different shielding configurations, as described in Ref.~\cite{Hanwook2020}. We compared heavy lead shielding with water shielding and ultimately decided to go for lead shielding due to the difficulty of implementing a cryostat in the water shielding. 
We compared the water Cherenkov detector with the plastic scintillation detector for the part above the cryogenic detector system. Based on simulation studies in Ref.~\cite{Jeewon2022}, we opted for a 70-cm-thick water Cherenkov detector with an active muon veto capability.
A thorough GEANT4 simulation was conducted to specify the thickness and layers of the various shielding materials. All of those parameters are reflected in the design of the AMoRE-II construction. 
As a result, we achieved a simulated background rate of 3.1$\times$10$^{-6}$ ckky in the (2.8--3.2) MeV energy region when the muon tagging efficiency with both PS and 70-cm-thick water tank is 93.9\%.

\subsection{Backgrounds from underground environments}
The experimental enclosure is surrounded by a rock wall finished with shotcrete. Due to the presence of $^{238}$U, $^{232}$Th, $^{235}$U, and $^{40}$K in the rock and shotcrete, this wall is a strong source of $\gamma$ rays. The rock sample, including shotcrete, is ground and measured by ICP-MS at the Korea Institute of Geoscience and Mineral Resource (KIGAM). The results of the activity measurements are utilized in this paper and will be published, along with measurements of the extensive detector and shielding materials for the AMoRE experiment, in a future publication.
Since the estimation of the heavy shielding materials requires significant CPU time, we simulate high-energy gammas from the decay of $^{214}$Bi larger than 3 MeV in order to estimate the background rate in the ROI.

Neutrons from ($\alpha$, n) natural radioactivity reactions in the rock and spontaneous fission, mainly of U atoms, in the rock are also possible background sources in underground environments. In particular, the thermal neutrons captured by copper or iron in the shield/structure materials generate high energy $\gamma$ rays around 7--8~MeV by (n,$\gamma$) reaction.
To evaluate the background effect due to these neutrons, we use the neutron flux measured with a Bonner sphere spectrometer system at Y2L~\cite{YSYoon2021}.
We simulate neutrons by generating them from the rock surface based on the measured energy spectrum.

In addition, as a noble gas, radon is chemically not very reactive and can diffuse easily through many materials infiltrating into the active region of the detectors~\cite{Next100:2016}.
One of the most significant isotopes of radon for background considerations is $^{222}$Rn, generated from the $^{238}$U decay chain in rock walls. Even though its half-life is relatively short ($T_{1/2}$=3.82 d), its decay products form the long-living radioisotope $^{210}$Pb ($T_{1/2}\approx 22$ yr). Moreover, several isotopes emitting high-energy gammas are fed by the subsequent decays of $^{222}$Rn, as stated in Ref.~\cite{Formaggio2004}.
Specifically, some emissions from decays $^{214}$Bi, a descendant of $^{222}$Rn decays, have energies over 3~MeV and can produce backgrounds in the 0$\nu\beta\beta$ ROI.
In order to minimize the effect of radon in the background, the Yemilab uses a radon reduction system (RRS) to supply air with radon levels reduced by at least 1000 times, resulting in  5~Bq/m$^{3}$, the estimated maximum Rn level. 
Additionally, a vinyl curtain surrounds the shielding structure and is flushed with air containing 5~Bq/m$^{3}$ radon. 
The radon concentration between the lead shield and cryostat is reduced by filling the space with urethane nitrogen-filled balloons. 
We simulated radon background events in the small amount of residual air between the OVC and the lead shielding to estimate the background rate in the ROI.
This simulation resulted in a rate of 1.7$\times$10$^{-5}$ ckky.

\subsection{Cosmogenic activation}
It is potentially dangerous for the AMoRE experiment if there is nuclear beta decay with a Q-value over 3 MeV inside crystals. This can be caused by activation of radioisotopes in the crystal detectors due to cosmic rays. 

At the AMoRE experimental site in the Yemilab, the rock overburden of approximately 1000 meters significantly reduces the muon flux. As a result, the effect of cosmogenic activation by muons and neutrons is negligible. However, during the crystal production process (including growing, polishing, cleaning, and gold deposition) at the Earth's surface and during transportation to the Yemilab, the crystals are exposed to cosmic rays.
This exposure may produce comparatively long-living radionuclides, contributing to the background.

To ensure the required radiopurity of the crystals, we used lithium carbonate and $^{100}$MoO$_{3}$ powders with a purity better than 99.998\%. We received about 180~kg of enriched molybdenum trioxide powder from Electrochemical Plant JSC~\cite{ecp_hp}. To prevent cosmogenic activation, the material was shipped to Korea by ground and sea. 
Upon arrival, the powder was stored in a desiccator at the Y2L, maintained at 23\textdegree C with a relative humidity of about 10\%. A 1L/min flow of boil-off gas from a liquid nitrogen dewar was used to optimize the storage conditions. The molybdenum powder was purified at the Center for Underground Physics (CUP), and purification efficiency and final product radiopurity were checked with the High Purity Germanium (HPGe) array at CUP (CAGe) and the Inductively Coupled Plasma Mass-Spectroscopy (ICP-MS)~\cite{SuyeonPark2019, SuyeonPark2020, HjYeon2023,Gileva2020,Gileva2017-qz}. Lithium carbonate precursor was used from two different sources. One is old stock preserved decades ago at the Nikolaev Institute of Inorganic Chemistry (NIIC), named NRMP TU 6-09-3728-83~\cite{KAShin2024} and produced at CUP powder~\cite{Olga2023,KAShin2024}.
 
We conducted simulations for all potential radionuclides meeting specific criteria, including origination by spallation from stable nuclides of Mo, Li, and O in the crystal, emission of beta/gamma-rays above 3 MeV, and having a half-live longer than 15 days.
We used the ACTIVIA~\cite{Back:2007kk} code to calculate radionuclide production rates for 30 days of exposure and 90 days of cooling underground. 

It found that the production of $^{82}$Sr (EC, 180~keV, 25.6~d)~/~$^{82}$Rb (EC, 4400~keV, 1.2~min) in Mo nuclides is most hazardous, with a background level of 10$^{-4}$ ckky. However, the half-life of $^{82}$Sr is only 25.6 days. $^{56}$Co (EC, 4566~keV, 77.3~d) in Mo is also dangerous, but most of these events are expected to occur within the first year. Hence, increasing the cooling time would help manage the risk associated with this radionuclide. The remaining cosmogenic nuclides are negligible, with a $\sim$10$^{-6}$ ckky background level.

Additionally, we considered the underground in-situ activation in detector modules, towers, cryostat, and shielding materials. Their impact on background contribution is negligible, below 10$^{-6}$ ckky.

\subsection{Backgrounds from the detector system} 
\label{sec:3.2}
\begin{table*}[!t]
\caption{Activities and fluxes of the background sources.} 
\label{table:activies}
\resizebox{0.95\textwidth}{!}{
\begin{tabular}{l l r r l l}
  \hline
  Material  & Supplier &  $^{238}$U($^{226}$Ra)  & $^{232}$Th($^{228}$Th) & $   $ Other & Technique\\
                       && (mBq/kg) & (mBq/kg) && \\
  \hline
  Araldite AW 106 CI & Huntsman           & 1.7(4)   & $<$ 1.0 && HPGe \\
  Araldite Hardner, HV953 U CI & Huntsman & 2.8(6)   & $<$ 1.2 && HPGe \\
  Si (heat detector wafer) & Microchemicals & $<$ 2    & $<$ 2 && HPGe \\
  Stycast 1266 resin & Loctite             & $<$ 1.1  & $<$ 1.2 && HPGe \\
  Stycast 1266 hardener & Loctite          & $<$ 1.1  & $<$ 3.1 && HPGe \\
  Pb/Sn solder (2023) & KNU               & $<$ 0.56 & $<$ 0.83 && HPGe \\
  Ultra-low Pb~\cite{CUORE2017} & Lemer Pax                & $<$ 0.05  & $<$ 0.05 && HPGe \\ 
  Pb brick & JL Goslar                    & 0.55(17) & 0.58(17) & $^{210}$Pb: 30(1) Bq/kg & CAGe \\
  Pb brick & Boliden                      & 0.48(12) & 0.45(11) & $^{210}$Pb: $<$ 10 Bq/kg& CAGe \\
  Pb brick & Haekgwang                    & 0.38(16) & $<$ 0.25   & $^{210}$Pb: $<$ 180 Bq/kg & CAGe \\
  STS 304 plate & POSCO                   & 1.00(16) & 2.36(22) && HPGe \\
  G11 & Leiden                            & 2700(200)& 906(66) && HPGe \\
  Urethane 0.3 mm & Seokyeong Industry    & $<$ 1.2  & $<$ 1.4 && HPGe \\
  Silicon & HRS Co.                       & $<$ 0.57 & 2.1(3) & $^{40}$K: $<$ 4.9 mBq/kg & HPGe \\
  Boric acid (99.99\%) & Alpha Aesar      & $<$ 0.46 & $<$ 0.50 & $^{40}$K: 98(8) mBq/kg & HPGe \\
  LMO crystal & CUP                       & 0.0020(3)&  0.0020(3) & $^{235}$U: 0.10(4) $\mu$Bq/kg & AMoRE-I \\

  \hline \hline
  Material  & Supplier &  $^{238}$U (pg/g)  & $^{232}$Th (pg/g) & $   $ Other & Technique\\
  \hline
  PTFE \cite{Aprile2017} & Maagtechnic            & $<$ 9.72 & $<$ 9.84 && ICP-MS \\
  Vikuiti film (roll type) & 3M                   & $<$ 3.6  & $<$ 4.5  && ICP-MS \\
  Polyimide-based, HGLS-D211EM & Hanwha L\&C      & 890(90)  & $<$ 1.2  && ICP-MS \\
  NOSV-Cu post & Aurubis (2021)                   & 0.38(4)  & 0.97(2)  && ICP-MS \\ 
  NOSV-Cu holder (top \& bottom) & Aurubis (2021) & 0.32(14) & 0.53(21) && ICP-MS \\
  OFE-Cu bulk & Aurubis (2021)                    & 0.83(11) & 0.98(14) && ICP-MS \\
  Brass screw & Sanco                    & 0.30(2) & 0.89(6) && ICP-MS \\ 
  \hline \hline
    \multicolumn{2}{l}{Background source} & \multicolumn{2}{c}{Flux} && \\
    \hline
    \multicolumn{2}{l}{Rock gamma}            & \multicolumn{2}{r}{$<$ 10.42 Bq/kg} &&\\
    \multicolumn{2}{l}{Radon air($^{222}$Rn)} & \multicolumn{2}{r}{$< 5^{\ast}$ Bq/m$^3$} &&\\
    \multicolumn{2}{l}{Cosmic muons \cite{Hanwook2020}}           & \multicolumn{2}{r}{$8.2 \times 10^{-8}$ muons/cm$^2$/s} && \\
    \multicolumn{2}{l}{Radiogenic neutrons \cite{YSYoon2021}}    & \multicolumn{2}{r}{7.1(10) $\times 10^{-6}$ counts/cm$^2$/s}&&\\
    \hline  
\end{tabular}}\\
{\footnotesize The values marked with asterisks ($\ast$) correspond to the AMoRE requirements.}
\end{table*}
\begin{table*}[!t]
  \caption{Simulated components.} 
  \label{table:simulated}
  \resizebox{0.95\textwidth}{!}{
  \begin{tabular}{l l l r}
      \hline
      AMoRE-II use & Simulated components & Material   & Simulated mass (kg) \\ 
      \hline
      Module &Heater adhesive              & Araldite (AW 106 CI:HV953 U CI = 1:1)  & 0.0002 \\ 
      &Heater                       & Si (heat detector wafer)                 & 0.0053 \\
      &Reflector                    & Vikuiti film (roll type)               & 0.38   \\
      &Clamps                       & PTFE                                   & 1      \\ 
      &Kapton PCB                   & Polyimide-based, HGLS-D211EM           & 0.09   \\ 
      &Solder for PCB               & Pb/Sn solder (2023)                    & 0.09   \\ 
      &Sensor adhesive              & Stycast 1266 (resin:hardener=100:28)   & 0.0002 \\
      &Phonon frame                 & NOSV-Cu holder                         & 26     \\ 
      &Photon frame                 & NOSV-Cu holder                         & 26     \\ 
      &Post (surface)                         & NOSV-Cu post                           & 23     \\ 
      &Screws for module            & Brass screw                   & 12     \\ 
      \hline
      Cryostat&Cu plate under the inner lead    & OFE-Cu bulk                    & 202.76 \\
      &Cooling plates supporting rod& G11                                    & 4.2\\ 
      &OVC                          & STS 304 plate & 599.13 \\ 
      &SC lead shield               & Ultra-low Pb (Lemer Pax)               & 51.48  \\
      \hline
      Shielding&Air balloon         & Urethane 0.3 mm                        & 6.84\\
      &Inner boric acid shield      & Boric acid (99.99\%)                   & 243    \\ 
      &Boric acid shield            & Silicon and boric acid (99.99\%)       & 1185    \\ 

      &Inner lead shield (1 cm)     & Ultra-low Pb (Lemer Pax)               & 86     \\
      &Inner lead shield (25 cm)    & Pb brick (JL Goslar)                   & 2055   \\ 
      &Lead shield (5 cm)           & Pb brick (Boliden)                     & 10865  \\
      &Lead shield (20 cm)          & Pb brick (Haekgwang)                   & 53839  \\ 
      \hline
  \end{tabular}}\\
\end{table*}
As detailed in Sec.~\ref{sec:experiment}, the detector system consists of crystal detector modules, cryostat, and shielding materials. 
Radiations originating from the LMO crystals are known to be the dominant source of backgrounds. 
For internal contaminations, radioactive $\alpha$-decay can be recognized by high-energy peaks in the background spectrum. However, individual $\alpha$ peaks resulting from the decays of $^{235}$U, $^{238}$U, and $^{232}$Th partially overlap in the spectrum. Thus, the internal radioactivity of LMO crystals is evaluated using $\alpha-\alpha$ time-correlated events, as described in Ref.~\cite{Alenkov2022}. 
In this study, we used the upper limit of internal activities obtained by analyzing $\alpha$ energy spectra measured cryogenically using LMO crystals in AMoRE-I at Y2L~\cite{Hanbeom2022}. 
These crystals were produced by CUP and NIIC using the same procedure employed in the production of the AMoRE-II crystals.

Other possible sources are activities from radioisotopes in the $^{238}$U, $^{232}$Th, $^{235}$U, and $^{40}$K decay chains from materials in the nearby crystal detectors, cryostat, and the shielding layers, which produce signals in the crystals. Therefore, this simulation considers all the materials used in the detector components and shielding layers.
These materials are scanned with either the HPGe detectors at the Y2L or by the ICP--MS equipment~\cite{Olga2023frontier}. Several materials are measured in both methods.
The activity measurements are listed in Table~\ref{table:activies} and are used to normalize the simulation results.

%% file: 4_analysis-results.tex
\section{Analysis and results}
\subsection{Event selection}
To construct the energy spectra of $\beta$/$\gamma$ background events in the simulations, we applied the same selection cuts used to background data processing in AMoRE-Pilot and AMoRE-I for the selection of 0$\nu\beta\beta$ decay event candidates. %
We consider $\beta$/$\gamma$ events with an assumption of almost 100\% $\alpha$ rejection power, based on a discrimination power (DP) of $\sim$10~$\sigma$ for energies around ROI~\cite{WTKim2022}. It is followed by single-hit selection, $\alpha$-tagging, and rejection of muon coincidence, which are itemized in the following list.

\begin{itemize}
  \item[$\bullet$]
We select single-hit $\beta$/$\gamma$ events classified as those with hits in only one of the crystals and none in any of the other crystals, considering a 5~ms time window, to reject background signals resulting from energy deposits in multiple crystals. \\

  \item[$\bullet$]
The $\beta$/$\gamma$ events from the decay of $^{208}$Tl in the $^{232}$Th chain can produce backgrounds in the ROI. However, they can be identified and rejected by using an $\alpha$-tagging method. This method involves checking for a time correlation with the $\alpha$ signal produced by the preceding $^{212}$Bi$\rightarrow$$^{208}$Tl $\alpha$ decay. We reject the events that occur within 30 minutes after an $\alpha$ event with 6207~$\pm$~50~keV in the same crystal. This results in a 98\% veto efficiency for beta events induced by $^{208}$Tl decay ($T_{1/2}$=3.05 min) in the crystals. \\

  \item[$\bullet$]
In order to avoid counting muon coincidence events, we tag muon events with an energy deposit above a set threshold in the plastic scintillator(s) and water tank.
Upon examining the time difference between veto hit and crystal hit from the simulation, we found that 99.5\% of muon-tagged events occur within 2 ms. 
Therefore, we set the veto time window to 5 ms to reject any events occurring within 5 ms (in this case the detection efficiency of the muon-tagged events increases to 99.7\%) following the appearance of a muon-tagged event in the muon veto detectors~\cite{Jeewon2022}. 
As a result, the estimated background rate of cosmic muons is significantly reduced from 1.65$\times$10$^{-3}$ ckky to 2.75$\times$10$^{-6}$ ckky by implementing muon detectors.
\end{itemize}

\subsection{Surface contamination}
The decay of surface $\alpha$ on crystals or nearby materials can deposit a range of energies, which can be as high as the Q-value of the decay. Sometimes, this energy falls in ROI. However, it can be distinguished from $\beta$/$\gamma$ signals through pulse shape
discrimination (PSD) and the light/heat ratio, 
with a DP of more than 14 at around 4.785~MeV~\cite{WTKim2022}. 
Nevertheless, events from the decay of radioactive contaminants in the crystal can pile up through subsequent daughter decay in the decay chains. When analyzing the real data, events that occur sequentially within approximately 2 ms for one crystal are considered indistinguishable. 
To assess the impact of such events on the background contribution, we carried out a study using beta-alpha pileup events considering a 5 ms time window. These $\beta$-$\alpha$ pileup events are from decays of ($^{212}$Bi+$^{212}$Po) in the $^{232}$Th decay chain, with a half-life of 294 ns, and decays of ($^{214}$Bi+$^{214}$Po) in the $^{238}$U decay chain, with a half-life of 164 $\mu$s. 
Our study found that these events contribute to the continuum down to the ROI region if they occur within the surface depth of the crystal, which is less than 50 $\mu$m. The estimated background level of these events is similar to that of $\beta$/$\gamma$ radiation of the crystal in ROI. However, the pileup events can be rejected by approximately 90\% within 0.5~ms (as discussed in Sect.~\ref{sec:4.3}), reducing them to approximately 10$^{-6}$ ckky, which is considered in this study.

In addition, we considered the nearby materials directly facing the crystals, like the copper holder, which houses the crystal and the wafer. The Copper frames and posts are made of NOSV copper, with each post featuring two screw threads. Despite undergoing a surface etching process, the machining of the screw threads may have introduced contamination that became deeply embedded in the posts, resulting in radioactivity levels more than ten times higher than for bulk NOSV copper~\cite{Olga2023frontier}. This contamination likely occurred during the thread-making process. Subsequently, we identified a company capable of producing copper posts with screw threads that exhibited significantly lower contamination levels, measuring 0.97 ppt for $^{232}$Th, as listed in Table~\ref{table:activies}. This is considered post-surface contamination, and its background contribution is included in our simulations, as reflected in Table~\ref{tab:result}, where the post-surface contribution is estimated to be 1.59 $\times$ 10$^{-5}$~ckky. 

\subsection{Background spectrum and rate in ROI}
\label{sec:4.3}
\begin{figure*}[ht]
\centering
\begin{tabular}{cc}
\includegraphics[width=0.49\textwidth]{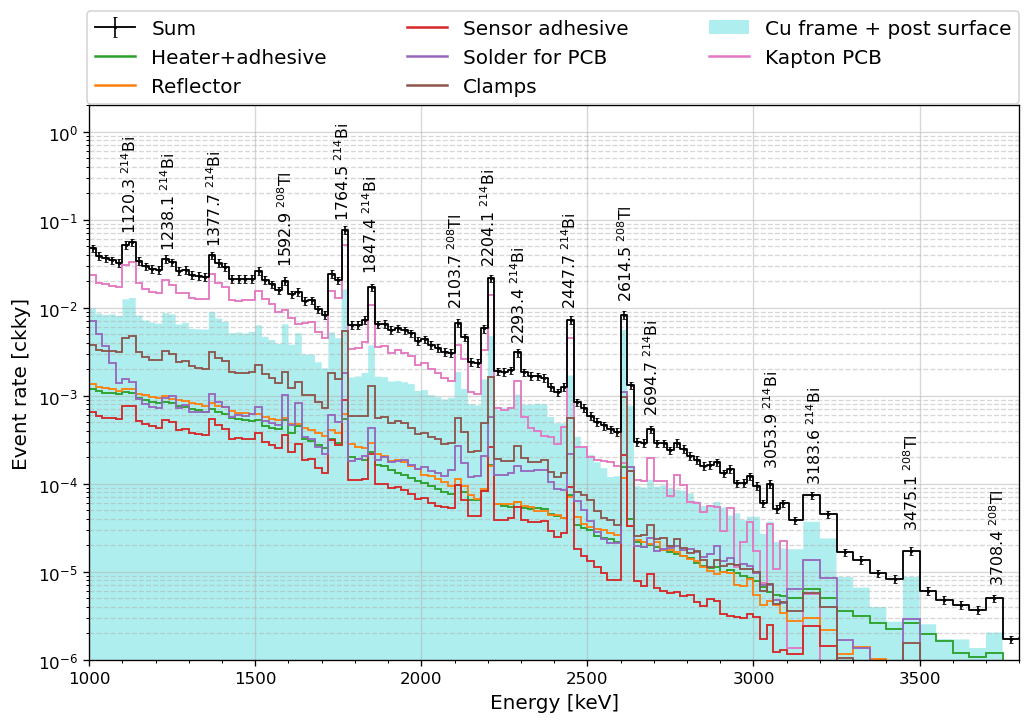} &
\includegraphics[width=0.49\textwidth]{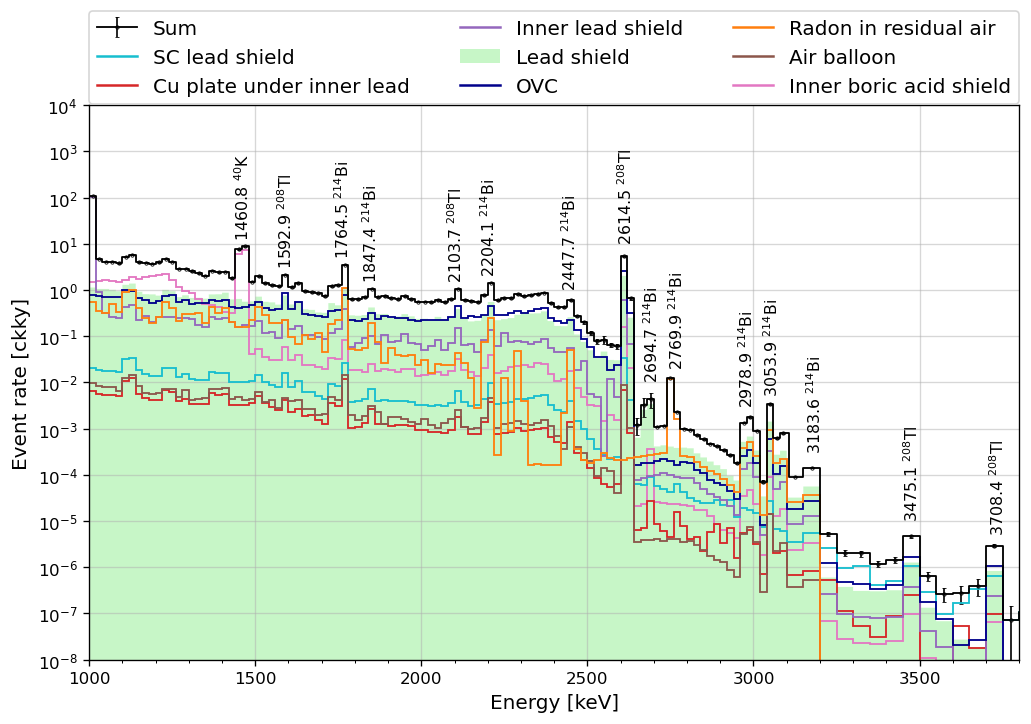} \\
(a) & (b) \\
\end{tabular}
\caption{Backgrounds from different components in the AMoRE-II setup: (a) G1: near-crystal components and (b) G2: far-crystal components}.
\label{fig:g1g2energysp}
\end{figure*}
\begin{figure*}[ht]
\centering
\includegraphics[width=0.8\textwidth]{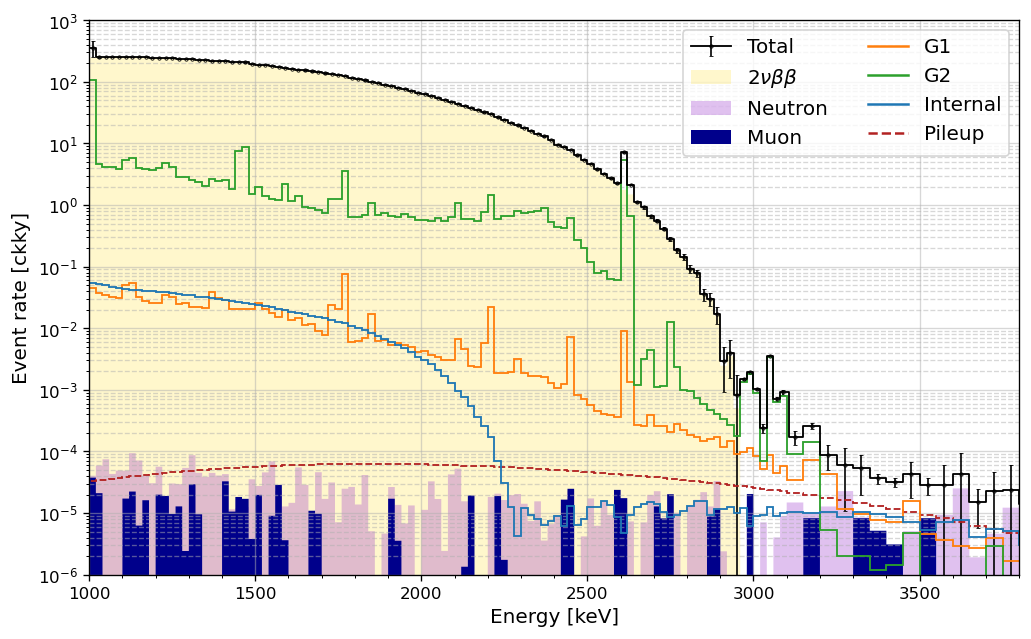}
\caption{Summed main components of the background spectrum after event selection}.
\label{fig:totalenergysp}
\end{figure*}
\begin{figure*}[ht]
\centering
\begin{tabular}{cc}
\includegraphics[width=0.5\textwidth]{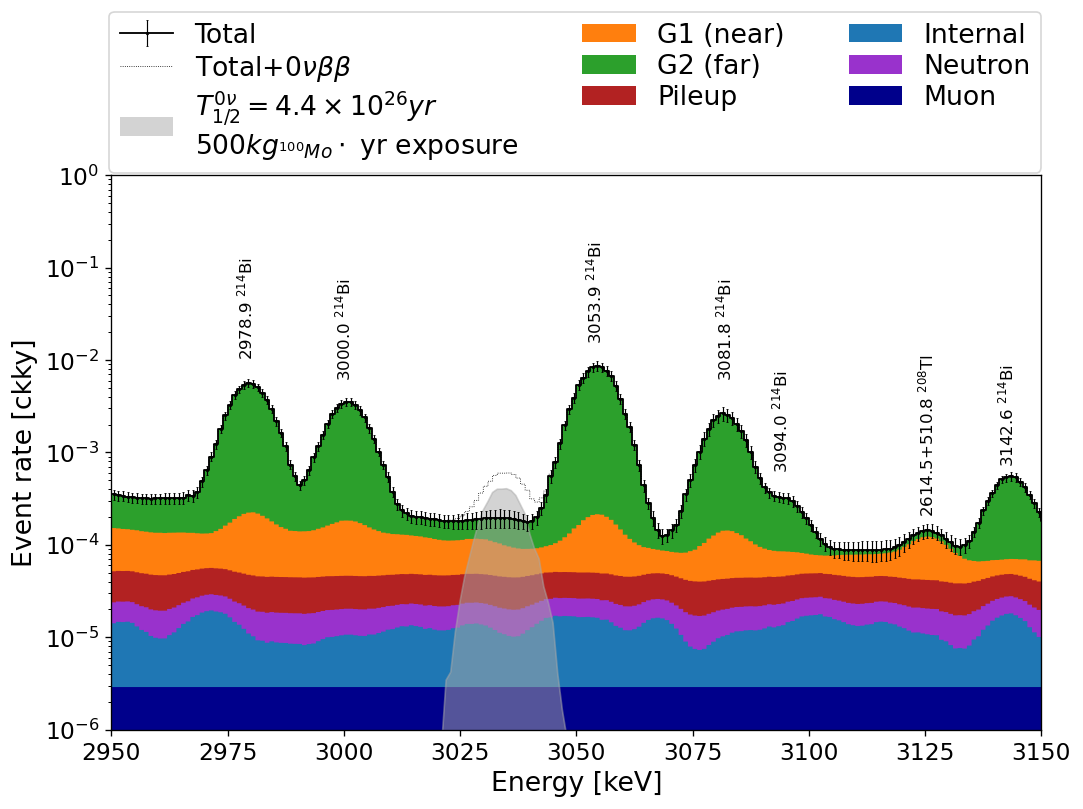} &
\includegraphics[width=0.47\textwidth]{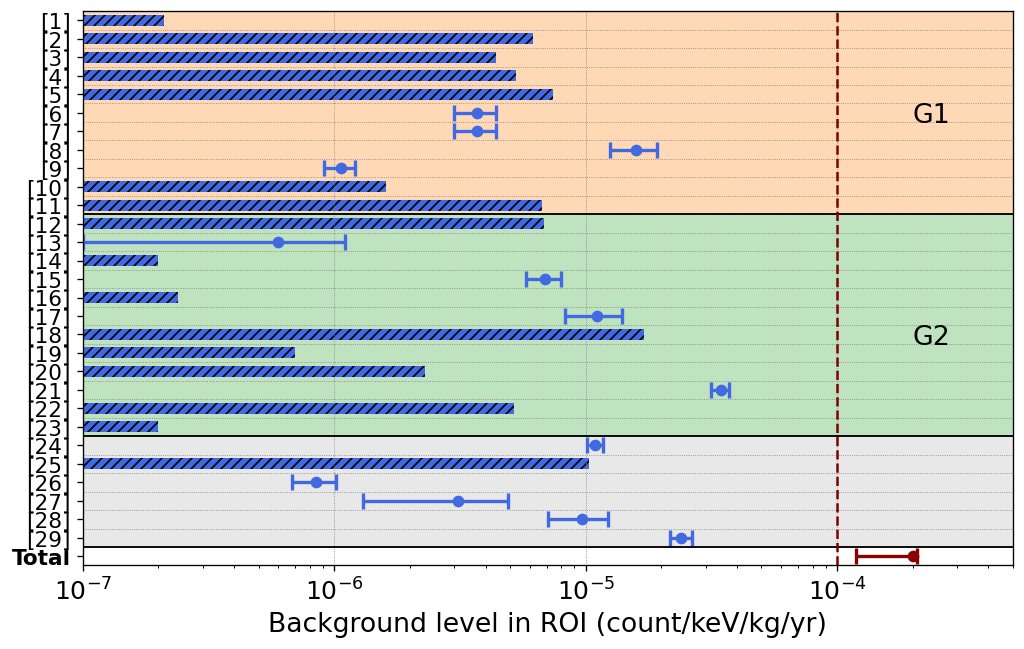} \\
(a) & (b) \\
\end{tabular}
\caption{Background spectrum in the extended ROI (a) and the rate in the ROI for the 29 components given in Table~\ref{tab:result} (b).}
\label{fig:roienergysp}
\end{figure*}
\begin{table*}[!h] 
\caption{Summary of the background event rates in the ROI for the major components.}
\centering
\label{tab:result}
\begin{tabular}[t]{ l r l r r r }

    \hline
    & No. & Components & \multicolumn{3}{c}{Background event rate ($\times 10^{-5}$ ckky)}\\[4pt] 
    &&& $^{238}$U($^{226}$Ra) & $^{232}$Th($^{228}$Th) & Total\\
    \hline
    Near component (G1) & 1 & Heater adhesive  & 0.006(1) & $<$ 0.015 &  $<$ 0.021 \\
    &2 & Heater             & $<$ 0.09 & $<$ 0.53 & $<$ 0.62 \\
    &3 & Reflector          & $<$ 0.20 & $<$ 0.24 & $<$ 0.44 \\
    &4 & Clamps             & $<$ 0.13 & $<$ 0.40 & $<$ 0.53 \\
    &5 & Kapton PCB         & 0.73(37) & $<$ 0.01 & $<$ 0.74 \\
    &6 & Phonon frame       & 0.05(4)  & 0.32(6)  & 0.37(7)  \\
    &7 & Photon frame       & 0.002(1) & 0.37(7)  & 0.37(7)  \\
    &8 & Post (surface)     & 0.29(18) & 1.30(29) & 1.59(34) \\
    &9 & Screws for module  & 0.005(4) & 0.101(15)  & 0.106(15) \\
    &10& Sensor adhesive    & $<$ 0.05 & $<$ 0.11 & $<$ 0.16 \\
    &11& Solder for PCB     & $<$ 0.03 & $<$ 0.64 & $<$ 0.67 \\    
    \hline
    Far component (G2) &12& SC lead shield  & $<$ 0.35 & $<$ 0.33 & $<$ 0.68 \\
    &13& Cu plate under the inner lead & 0.05(4)     & 0.010(8) & 0.06(5) \\
    &14& Inner lead shield (1 cm)   & $<$ 0.021 & $<$ 0.0003 & $<$ 0.02 \\
    &15& Inner lead shield (25 cm)   & 0.68(12)  & 0.013(1) & 0.69(12) \\ 
    &16& Cooling plates supporting rod & $<$ 0.02$^{\ast}$& $<$ 0.004$^{\ast}$ & $<0.024^{\ast}$\\ 
    &17& OVC                       & 1.05(26)        & 0.058(6)         & 1.11(28) \\
    &18& Radon in residual air     & & & $<$ 1.7 \\
    &19& Air balloon               & $<$ 0.07        & $<$ 0.003        & $<$ 0.04 \\
    &20& Inner boric acid shield   & $<$ 0.23        & $<$ 0.003        & $<$ 0.23 \\
    &21& Lead shield (5 cm)        & 3.40(28)        & 0.039(3)         & 3.44(28) \\
    &22& Lead shield (20 cm)       & 0.52(4)         & $<$ 0.002        & $<$ 0.52 \\ 
    &23& Boric acid shield         & $< 0.015^{\ast}$ & $< 0.002^{\ast}$& $< 0.017^{\ast}$ \\
    \hline
    &24&LMO (internal)        & 0.08(2) & 1.01(7) & 1.09(8) \\
    &25&$\gamma$ from rock    & & & $<$ 1.03 \\
    &26&Solar $\nu$           & & & 0.085(17) \\
    &27&Cosmic muons           & & & 0.31(18) \\
    &28&Radiogenic neutrons    & & & 0.97(26) \\
    &29&$2\nu\beta\beta$ random coincidence  & & & 2.40(24) \\
    \hline
    Total &&                & & & $20.06^{+ 0.68}_{- 8.24}$ \\
    \hline
\end{tabular}\\
{\footnotesize ($\ast$) represents the 90\% confidence level (C.L.) with zero events in the ROI.}
\end{table*}
Figure~\ref{fig:g1g2energysp} shows simulated energy spectra with different colors for all possible background sources grouped in categories, 
which we obtained after applying event selection cuts to the simulated events and convolving them with energy resolution as a function of energy.  
We categorized the background sources from the detector system into two groups: components located near and far from the crystal. 
The near-crystal components (G1) include Araldite, bolts, clamps, Stycast, Kapton PCB, heaters, Pb-Sn solder, PTFE, Vikuiti reflector film, and copper materials used in the phonon/photon frame. The far-crystal components (G2) comprise OVC, Radon, SC-shield, lead shield, and other shielding materials.

In addition, there are other groups of background sources such as crystal internal background, 2$\nu\beta\beta$ decay, pileup events, neutrons, muons, and solar neutrinos.
The background from interactions of solar neutrinos with the $^{100}$Mo target is estimated to be 8.5$\times$10$^{-7}$ ckky at the energy of 3034~keV~\cite{Bahcall2001,Ejiri2017}. This is mainly from single $\beta$ decay of $^{100}$Tc.
Electrons emitted in $\nu_{e}$ capture itself, and scattering of $\nu_{e}$, $\nu_{\mu}$, $\nu_{\tau}$ on electrons give much
smaller contributions.

The energy spectrum summed over all simulations is shown in Fig.~\ref{fig:totalenergysp} by a solid black line with 1~$\sigma$ error bars. The background contribution below 3 MeV is mainly from 2$\nu\beta\beta$ decay, which is expected due to the use of purified materials.

The high-energy spectrum in the extended ROI ranging from 2.95 to 3.15~MeV is shown in Fig.~\ref{fig:roienergysp} (a), where the main contributions are from the background sources of $^{214}$Bi in the $^{226}$Ra--$^{210}$Pb decay sub-chain of $^{238}$U and $^{208}$Tl in the $^{228}$Th--$^{208}$Pb decay sub-chain of $^{232}$Th.

The peaks at 2978.9, 3000., 3053.9, 3081.8, and 3142.6~keV are attributed to high-energy $\gamma$ emissions from the decay of  $^{214}$Bi in far-crystal components (G2), such as a 5-cm thick Boliden lead shield layer, OVC, and radon in the air between them.

To eliminate $^{208}$Tl decay events, a vetoing process is employed 30 minutes after the 6.2 MeV $\alpha$ precursor as the $\alpha$-tagging method.
However, when dealing with backgrounds from $^{208}$Tl decay in materials located near-crystal components (G1), such as Pb-Sn solder and Kapton PCB, the precursor $\alpha$ may not provide an energy level of 6.2~MeV. It can lower veto efficiency and result in background contributions in ROI.

In addition, another background contribution to the ROI is pileup events due to the random coincidence of two $^{100}$Mo 2$\nu\beta\beta$ decay events. 
We investigated the rate of this background using the DECAY0 program~\cite{Ponkratenko2000} to produce $^{100}$Mo 2$\nu\beta\beta$ decay events. The estimated background rate within the ROI, which occurred randomly within a 1 ms time window, was 1.2$\times$$10^{-4}$ ckky for 310~g of $^{40}$Ca$^{100}$MoO$_{4}$~\cite{Luqman2017}. Furthermore, we found that a single 2$\nu\beta\beta$ decay that coincides randomly with other background sources has an upper limit of 1.1$\times$$10^{-4}$ ckky in ROI.
To reduce the background level, we studied rejection efficiency for pileup events~\cite{GBKim2017} and further developed it to meet the AMoRE-II background requirement in a separate work by utilizing various PSD parameters as input parameters for the Boosted Decision Tree (BDT) method.
We found that it is possible to reject approximately 90\% of pileup events within 0.5 ms, resulting in a background rate of 2.4$\times$$10^{-5}$ ckky for a 6 cm diameter LMO crystal, which includes a single 2$\nu\beta\beta$ decay that randomly coincides with other background sources. This information is used in this paper.

The background contribution in the (3024--3044)~keV range (ROI) of each background source included in the group is shown in Fig.~\ref{fig:roienergysp} (b). 
A bar graph represents the background rate based on a conservative detection limit for the impurity. Meanwhile, the filled circle with error bars is estimated based on the impurity measurement.
One of the primary sources of background radioactivity is $^{214}$Bi, which is found in the $^{226}$Ra--$^{210}$Pb decay sub-chain of $^{238}$U located in the lead shielding's innermost layer. %
The inner 5~cm of current shielding with Boliden lead will be replaced by lead with a lower $^{226}$Ra contamination. This replacement aims to reduce the background level to approximately $10^{-5}$~ckky, which can be achieved using a purer lead with a radiopurity of less than 200~$\mu$Bq/kg for $^{226}$Ra.

The detailed results estimated in the ROI of the radioactive sources considered in this study for the background contributions from radioisotopes in the decay chains of the $^{238}$U and $^{232}$Th are summarized in Table~\ref{tab:result}. 
The summary table also includes evaluations of other background groups such as internal, 2$\nu\beta\beta$, neutrons, muons, and solar neutrinos.
The total estimated background level is 2.01$\times$$10^{-4}$~ckky.
Additionally, the study assessed the background effects of other radioisotopes, such as $^{235}$U, $^{40}$K, and $^{210}$Pb, and found that they have negligible contributions.

%% file: 5_sensitivity.tex
\section{Estimation of half-life sensitivity}
\begin{figure}[t]
\centering
\includegraphics[width=0.48\textwidth]{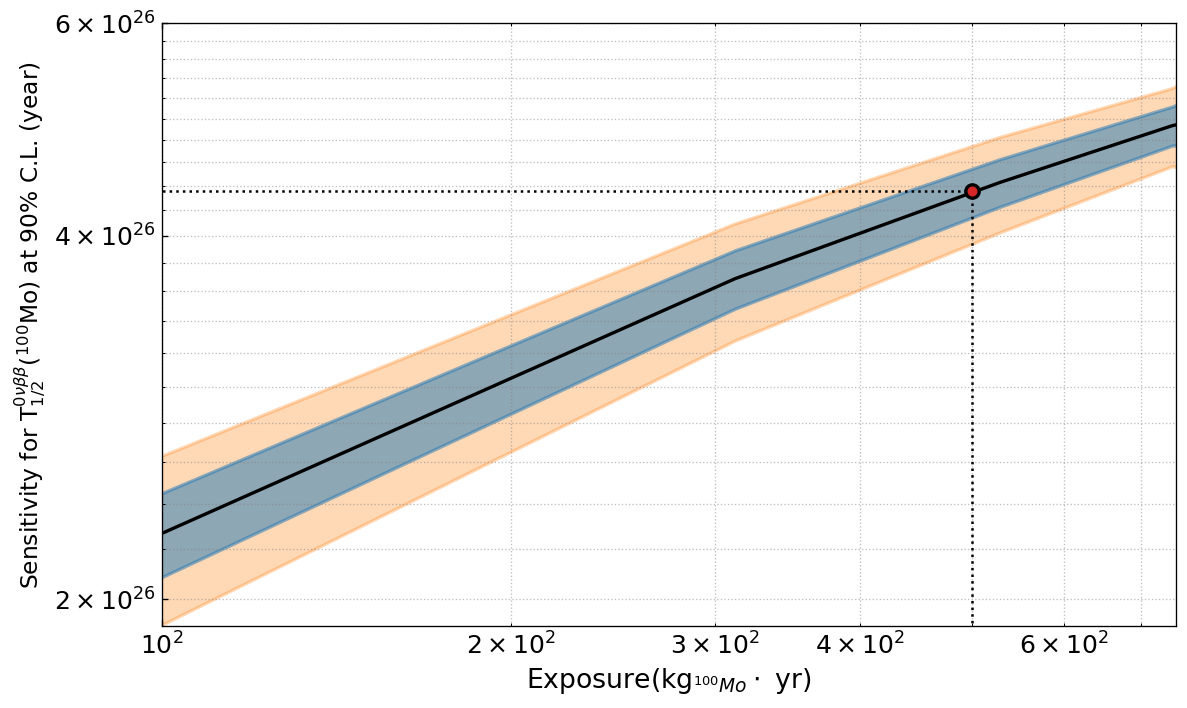}
\caption{Median sensitivity of $\mathnormal{T}$$_{1/2}^{0\nu\beta\beta}$ as a function of exposure in units of $\text{kg}\cdot \text{year}$, with $\pm$1 $\sigma$ ($\pm$2 $\sigma$) error band shown in blue (orange).}
\label{fig:sensitivity}
\end{figure}
To study the sensitivity of the half-life of $\mathnormal{T}$$^{0\nu\beta\beta}_{1/2}$, taking into account the background spectra estimated in Sect.~\ref{sec:4.3}, we generate $10^{4}$ pseudo-experiments that are analyzed using a fitting procedure as it is done for the measured data. %
For a single pseudo-experiment for the null hypothesis, the event rate over energy in a single-hit energy spectrum is generated based on Poissonian fluctuations in each 1~keV energy bin of the background spectrum for each background source estimated in Sect.~\ref{sec:4.3}. 
 
We use the signal spectrum of $0\nu\beta\beta$ decay simulated and convolved by the energy-dependent resolution and the background spectrum estimated to fit the pseudo-experiment data by using a log-likelihood method.
The likelihood function, $\mathcal{L}$, is given by a binned Poisson likelihood: 
\begin{align}
	\mathcal{L} 	&=
	 	\begin{aligned}[t]
		&\prod_{i}^{N_{bins}}{\frac{(\mu_{i})^{n_{i}}}{n_{i}!}e^{-(\mu_{i})}} 
		\end{aligned}
\end{align}		
where $n_{i}$ represents the number of observed events in the $i^{th}$ energy bin of a pseudo-experiment data and $\mu_{i}$ represents the number of expected events at the same energy bin.  $\mu_{i}$ can be obtained by using a model function defined as:
\begin{align}
	\mathcal{\mu}_{i}	&= 
	 	\begin{aligned}[t]
  		&S\cdot \mathcal{P}_{s}(i) + \sum_{j} B_j \cdot \mathcal{P}_{b_j}(i)
		\end{aligned}
\end{align}		
where $\mathcal{P}_{s}(i)$ and $\mathcal{P}_{b_j}(i)$ are the probability density of the models for the signal and the $j^{th}$ background component spectra, respectively, at the $i^{th}$ energy bin, $S$ and $B_{j}$ represent the number of events in the fitting range of [2950, 3150]~keV of the signal and the $j^{th}$ background component spectra.
$\mathnormal{S}$ is expressed in terms of the decay rate of $0\nu\beta\beta$ decay, denoted by $\Gamma^{0\nu}$ as below:
\begin{align}
	S &=
		\epsilon \cdot \Gamma^{0\nu} \cdot N_{^{100}\text{Mo}} \cdot \Delta{t} \\
		&=
		\epsilon \cdot \frac{\ln{2}}{T^{0\nu\beta\beta}_{1/2}} \cdot \frac{N_{A}}{A_{\text{LMO}}} \cdot a \cdot \mathcal{M} \cdot \Delta{t}
\end{align}
where $\epsilon$ is a detection efficiency of 0.7~\cite{Alenkov2019}, ${N_{A}}$ is the Avogadro’s number, $A_{\text{LMO}}$ is the molar mass of Li$_{2}$$^{100}$MoO$_{4}$, $a$ is the concentration of $^{100}$Mo, $\mathcal{M}$ is the total detector mass, and $\Delta{t}$ is the exposure time.
In the fit procedure, $\Gamma^{0\nu}$ and $B_{j}$ are considered fit parameters. 

We obtain the exclusion limit for the half-life of $0\nu\beta\beta$ decay at a 90\% confidence level (C.L.) by fitting the model spectrum to pseudo-experiment data. To build the distribution of the half-life exlusion limits at a 90\% C.L. as a function of exposure in units of \text{kg}$_{^{100}Mo}\cdot \text{year}$, we generated $10^{4}$ pseudo-experiments with a step size of 0.1 years for ten years of exposure time. The median of the distribution of 90\% C.L. half-life exclusion limits is shown as a function of exposure in Fig.~\ref{fig:sensitivity}. %
The blue and orange colors represent the 1 and 2 sigma bands of the median sensitivity, respectively.
The half-life exclusion limit at 90\% C.L. for AMoRE-II with an exposure of 500~\text{kg}$_{^{100}Mo} \cdot$ \text{year} is estimated to be 4.4~$\times$~10$^{26}$ years, corresponding to the effective Majorana mass of (18~--~54)~meV. To improve this limit, it is essential to reduce the background level, which can be achieved by using higher purity materials and lowering impurity detection limits in radioassay campaigns of the materials. Reducing the background levels will ultimately enhance the experimental sensitivity.

We compared a simulated signal of $0\nu\beta\beta$ decay of $^{100}$Mo in the target crystal with the predicted background spectrum using a detector efficiency of 0.7 and an exposure time of 5.2 years. The simulated signal is normalized by the activity corresponding to $\mathnormal{T}^{0\nu\beta\beta}_{1/2}$ of 4.4~$\times$~10$^{26}$ years. The results are shown in Figure~\ref{fig:roienergysp}~(a). The dotted grey line represents the sum of the background and signal. We expect this sum to be observed by AMoRE-II after approximately five years of exposure.

%% file: 6_conclusion.tex
\section{Conclusions}
AMoRE-II uses a 180~kg mass of 423 LMO crystals to improve the limit on the half-life of $0\nu\beta\beta$ decay from $^{100}$Mo. 
In order to achieve the background level of 10$^{-4}$~ckky, Monte-Carlo simulations based on the Geant4 toolkit are intensively conducted. This results in a background level of 2.0~$\times$~10$^{-4}$~ckky in the ROI. The main source of backgrounds is $^{214}$Bi in the $^{226}$Ra--$^{210}$Pb decay sub-chain of $^{238}$U located in the innermost layer of the lead shielding.
To meet the goal of 1~$\times$~10$^{-4}$~ckky, further reduction of the setup materials radioactivity may be necessary. This could be achieved by substituting the current shielding material with higher-purity lead and lowering impurity detection sensitivity in radioassay campaigns of the materials.
Taking into account the projected background spectrum and signals from $0\nu\beta\beta$ decay, we estimated the median sensitivity of $^{100}$Mo's $0\nu\beta\beta$ decay half-life limits at a 90\% confidence level using pseudo-experiment data.
The estimated sensitivity is (4.4~$\pm$~0.2)$\times$10$^{26}$ years ((4.8~$\pm$~0.2)$\times$10$^{26}$ years) with a background level of 2.0~$\times$~10$^{-4}$~ckky (1.0~$\times$~10$^{-4}$~ckky) for an exposure of 500~\text{kg}$_{^{100}Mo} \cdot$ \text{year}.